\documentclass[aps,pra, twocolumn]{revtex4}
\usepackage{graphicx}    
\usepackage{amsmath}
\begin{document}

\title{Quantum entropy and polarization measurements of the two-photon system}
\author{\textbf{Moorad Alexanian}$^{1}$ and \textbf{Vanik E. Mkrtchian}$^{2}$
\\
$^{1}$\textit{Department of Physics and Physical Oceanography University of
North Carolina Wilmington, NC 28403-5606}\\
[1ex] $^{2}$\textit{Institute for Physical Research, Armenian Academy of
Sciences. Ashtarak, 0203, Republic of Armenia} \\
[1ex] }

\date{\today}
\begin{abstract}
\noindent We consider the bipartite state of a two-photon polarization system and obtain the exact analytical expression for the von Neumann entropy in the particular case of a 5-parameter polarization density matrix. We investigate and graphically illustrate the dependence of the entropy on these five parameters, in particular, the existence of exotic, transition from exotic to non-exotic, and non-exotic states, where the quantum conditional entropy is negative, both positive and negative, and positive, respectively. We study the ``cooling" or ``heating" effect that follows from the reduced density of photon 2 when a measurement is performed on photon 1.

\end{abstract}
\maketitle
\section{Introduction}

The philosophical implications of the superposition principle of quantum mechanics and quantum inseparability have profound effects for our understanding of the physical aspect of nature, which differs drastically from that provided by classical mechanics \cite{AP95}. Quantum entanglement or quantum inseparability is essential for quantum information communication and processing protocols in quantum cryptography \cite{E91}, dense coding \cite{BW92}, teleportation \cite{BBC93}, and entanglement swapping \cite{PBW98}, which can be used to realize quantum repeaters \cite{BDC98}. Entanglement can be achieved via two interacting quantum systems \cite{KMW95} or by an appropriate joint measurement of two systems \cite{W94}.

In handling quantum information, the notion of quantum noise or decoherence enters as an unwanted interaction with the outside world. Here we deal with the dynamics of a closed quantum system and study the polarization of a two-photon system and measurements performed on the system. We show the existence of exotic states and their transition into non-exotic state (and vice versa) by suitably tuning the parameters, which is interesting in some aspects of quantum computation and quantum information. Several authors have studied how to protect coherence and thus control dissipation \cite{GLL06, OLJ15, AVV17,DKD18,CKS18}.

The role of the von Neumann entropy \cite{Neum} has gained in importance in recent years owing to the extensive development of the physics of entangled quantum states. The entropy plays a fundamental role as a quantitative measure of entanglement \cite{Woot, RMP}. Accordingly, any exact result of the von Neumann entropy for a correlated quantum system is of extreme importance. In this paper, we consider the bipartite system of a pair of photons and exactly calculate the general expression for the von Neumann entropy for correlated, polarized states \cite{Fano,vem}. The generalized conditional entropy has been analyzed for this system \cite{GR14} for which knowledge of the eigenvalues of the density matrix is not required.

Correlation transfer from one-photon to two-photon systems, not in any restricted subspace, but in the complete space of the polarization degree of freedom has been studied \cite{KSJ16}.  Three protocols for directly measuring the concurrence of two-photon polarization-entangled states, including pure states and mixed states has been considered \cite{ZYY13}. An experimentally realizable scheme for manipulating the entanglement of an arbitrary state of two polarization-entangled qubits has been introduced, where the von Neumann entropy provides a convenient and useful measure of the purity of the state \cite{TM01}.

We investigate the nature of the reduced density matrix of photon 2 after a measurement is performed on photon 1. This measurement gives rise to a ``cooling" (``heating") effect of photon 2 if the final entropy of photon 2 is less (greater) than the single photon entropy.

This paper is arranged as follows. In Sec. II, we review the conditional quantum entropy used in quantum information theory. In Sec. III, we consider the general expression for the von Neumann entropy for two photons. In Sec. IV, we study the four eigenvalues for the two-photon polarization density matrix. In Sec. V, we study two-photon correlations for a 5-parameter family of density matrices and obtain results for the entropy for three models that illustrate a wide-ranging behavior of the von Neumann entropy. In particular, the transition from negative to positive values of the quantum conditional entropy as a function of the photon Stokes parameter, viz., the transition from exotic to non-exotic states, as well as, the existence of strictly exotic and non-exotic states. In Sec. VI, we consider a measurement of one of the photons with an analyzer and the corresponding effect that this measurement has on the reduced entropy of the second photon. Finally, Sec. VII summarizes our results.

\section{Quantum conditional entropy}

The quantum conditional entropy is defined by \cite{NC2000}
\begin{equation}
S(\hat{\rho}^{(1)}|\hat{\rho}^{(2)})=S(\hat{\rho}^{(1,2)})-S(\hat{\rho}^{(2)})
\end{equation}
while the quantum mutual information by
\begin{equation}
S(\hat{\rho}^{(1)}:\hat{\rho}^{(2)})= S(\hat{\rho}^{(1)}) + S(\hat{\rho}^{(2)})-S(\hat{\rho}^{(1,2)}),
\end{equation}
where $S(\hat{\rho}^{(1,2)})$ is the joint entropy for the composite system and $S(\hat{\rho}^{(i)})$ is the entropy of the $i$-th component, $i=1,2$. Henceforth, $S\equiv S(\hat{\rho}^{(1,2)})$ and $S_{i}\equiv S(\hat{\rho}^{(i)})$. For entangled states, one has the entropic inequality \cite{HHHH09}
\begin{equation}
S< S_{i}.
\end{equation}
However, there are entangled states which do not exhibit this exotic property (3) but rather satisfy the inequality $S \geq S_{i}$

The conditional quantum entropy is an entropy measure used in quantum information theory. An important feature of the quantum conditional entropy (1) is that it can assume negative values \cite{CA97}, that is, the subsystems of the entangled system can exhibit more disorder than the system as a whole \cite{HH1994}. The negativity of quantum conditional entropy is a sufficient criterion for quantum non-separability and gives the additional number of bits above the classical limit that can be transmitted in a quantum dense coding protocol.

The conditional entropy with a minus sign is known as the coherent information and is a fundamental quantity responsible for the capability of transmission of quantum information. The classical analog of the conditional entropy indicates that the Shannon entropy of a single random variable is never larger than the Shannon entropy of two variables \cite{HHHH09}. Accordingly, a negative quantum conditional entropy is a definite signature of quantum-entangled states.

\section{Von Neumann entropy for two-photon system}

The general form of the normalized polarization density matrix for two
photons \cite{Fano,vem} is given by%
\[
\hat{\rho}^{(1,2)}=(1/4)\Big{(} \hat{I}^{( 1)}\otimes \hat{I}^{( 2) }+ \hat{\boldsymbol{\xi}}^{(1)}\cdot\hat{\boldsymbol{\sigma}}^{(1)}\otimes \hat{I}^{( 2) }
\]
\begin{equation}
+\hat{I}^{( 1)}\otimes \hat{\boldsymbol{\xi}}^{(2)}\cdot\hat{\boldsymbol{\sigma}}^{(2)}+\sum_{i,j}\zeta _{ij}\hat{\sigma}_{i}^{( 1)}\otimes \hat{\sigma}_{j}^{( 2)}\Big{)},
\end{equation}
where $\hat{I}^{\left( 1\right) },\hat{I}^{\left( 2\right) }$ and $\boldsymbol{
\hat{\mathbf{\sigma}}}^{\left( 1\right) },\boldsymbol{\hat{\sigma}}^{\left( 2\right) }$
are $2\times 2$ Pauli vector matrices acting in the polarization space of photons
and the real dimensionless quantities $\boldsymbol{\xi }^{\left( 1\right) }\boldsymbol{,\xi }%
^{\left( 2\right) },\zeta _{ij}$ $\left( i,j=1,2,3\right) $ are functions of the
photon momenta and of the parameters of the emitting system. The dimensionless vectors $\boldsymbol{\xi }^{\left(
1,2\right) }$ are the Stokes vectors of photons $1,2$, respectively, while the parameter $\zeta _{ij}$
describes the two photon polarization correlation. In the case of no photon entanglement, one has that %
\begin{equation}
\zeta _{ij}=\xi _{i}^{\left( 1\right) }\xi _{j}^{\left( 2\right) }.
\label{eq:(2)}
\end{equation}%
Generally, the Stokes parameters $\hat{\boldsymbol{\xi }}^{\left(
1,2\right) }$ and $\zeta _{ij}$ satisfy the inequalities \cite{vem}%
\begin{equation}
\left\vert \xi _{i}^{\left( 1\right) }+\xi _{j}^{\left( 2\right)
}\right\vert -1\leq \zeta _{ij}\leq \left\vert \xi _{i}^{\left( 1\right)
}-\xi _{j}^{\left( 2\right) }\right\vert +1  \label{eq:(3.a)},
\end{equation}%
\begin{equation}
\hat{\boldsymbol{\xi}}^{(1)}\cdot\hat{\boldsymbol{\xi}}^{(1)}+\hat{\boldsymbol{\xi}}^{(2)}\cdot\hat{\boldsymbol{\xi}}^{(2)}+\sum_{i,j}\zeta
_{ij}^{2}\leq 3  \label{eq:(3.b)}.
\end{equation}%

The reduced density matrix for photon $1$, viz., $\hat{\rho}^{\left( 1\right) }$, is obtained by taking the trace of (4) over the quantum states of photon $2$, which gives us the Stokes matrix of photon $1$,
\begin{equation}
\hat{\rho}^{\left( 1\right) }=Tr_{2}\hat{\rho}^{\left( 1,2\right) }=(1/2)[ \hat{I}^{\left( 1\right) }+ \hat{\boldsymbol{\xi}}^{(1)}  \cdot \hat{\boldsymbol{\sigma}}^{(1)}].
\end{equation}
Similarly, by taking a trace of (4) over the quantum states of photon 1, one obtains the Stokes matrix of photon $2$, viz., $\hat{\rho}^{\left( 2\right) }$,
\begin{equation}
\hat{\rho}^{\left( 2\right) }=Tr_{1}\hat{\rho}^{\left( 1,2\right) }=(1/2)[ \hat{I}^{\left( 2\right) }+\hat{\boldsymbol{\xi}}^{(2)}\cdot\hat{\boldsymbol{\sigma}}^{(2)}] .  \label{eq:(4.b)}
\end{equation}

The goal of our paper is to calculate the von Neumann entropy \cite{Neum}%
\begin{equation}
S(\hat{\rho}^{(1,2)})=-Tr(\hat{\rho}^{\left( 1,2\right) }\ln \hat{\rho}^{\left( 1,2\right) })
\label{eq:(5)}
\end{equation}%
of the two photon system in a mixed quantum state, that is, when%
\begin{equation*}
\hat{\rho}^{2}\neq \hat{\rho}.
\end{equation*}%
Araki and Lieb \cite{Araki} have proven that %
\begin{equation}
\left\vert S_{1}-S_{2}\right\vert \leq S\leq S_{1}+S_{2},  \label{eq:(6.a)}
\end{equation}%
where $S_{1},S_{2}$ are the reduced von Neumann entropies of photons $\alpha=1,2$
\begin{equation}
S_{\alpha}=\ln 2-(1/2)\ln [(1-\xi ^{\left( \alpha \right) })^{1-\xi
^{\left( \alpha \right) }}(1+\xi ^{\left( \alpha \right) })^{1+\xi ^{\left(
\alpha \right) }}],
\end{equation}
where $\xi^{(\alpha)}$ is the magnitude of the Stokes vector $\hat{\boldsymbol{\xi}}^{(\alpha)}$.

The entropy $S_{\alpha}$ is a monotonically decreasing function of the Stokes
parameter for $0 \leq \xi^{(\alpha)}\leq 1 $ and achieves its maximum value $\ln 2$ for the completely unpolarized state when $\xi^{(\alpha)} =0$
and its minimum value zero when the photon is in a pure polarized state, viz., $\xi^{(\alpha)} =1.$

\section{Entropy for two-photon polarization density}

The arbitrary polarization state of the photon pair (4) is described by 15 real
parameters \cite{vem} and owing to the Araki-Lieb inequality (11), the maximum entropy attainable is%
\begin{equation}
S_{\max }=2\ln 2  \label{eq:(7)}
\end{equation}%
when all of 15 parameters are set equal to zero. That is to say, no correlations of any kind, neither individually nor for the pair of photons.

The eigenvalue equation of the matrix (4) is
\begin{equation}
\lambda ^{4}-\lambda ^{3}+c_{2}\lambda ^{2}-c_{1}\lambda +c_{0}=0,
\label{eq:(8.a)}
\end{equation}%
where the coefficients are defined as follows
\begin{equation*}
c_{2}=\left( 1/8\right) p,
\end{equation*}%
\begin{equation}
c_{1}=\left( 1/16\right) \left[ p-2(1-\sum_{i,j}\xi _{i}^{\left( 1\right) }\zeta
_{ij}\xi _{j}^{\left( 2\right) }+\det \hat{\zeta})\right],  \label{eq:(8.b)}
\end{equation}%
\begin{equation*}
c_{0}=\det \hat{\rho}^{\left( 1,2\right) }.
\end{equation*}
In (15), $p$ is a nonnegative number (see (7)) which
is called the \textit{purity} of the state \cite{Gem}
\begin{equation}
p=3-\hat{\boldsymbol{\xi}}^{(1)}\cdot\hat{\boldsymbol{\xi}}^{(1)}-\hat{\boldsymbol{\xi}}^{(2)}\cdot\hat{\boldsymbol{\xi}}^{(2)}-\sum_{i,j}\zeta _{ij}^{2}  \label{eq:(9)}
\end{equation}%
and describes the ``distance" of the mixed state of the system from the pure state where $p=0$. \cite{Fano,vem}%

The four non-negative solutions $\lambda _{j}\left( j=1,..4\right)$ of (14) yield the von Neumann entropy%
\begin{equation}
S=-\sum_{j}\lambda _{j}\ln \lambda _{j}.  \label{eq:(10)}
\end{equation}
The roots of the quartic equation (14), with coefficients given by (15), are extremely awkward to write and thus the exact expression for the entropy (17) would not be too useful. Accordingly, we consider special cases of (4) that lead to the quartic form (14) being reduced to the product of two quadratic polynomials

\section{Photon correlations}

We consider the special case of the density matrix (4) which contains the
main properties of the two-photon system and which is easier to handle mathematically than the general case that includes all 15 real parameters. Namely, we consider the
case where the 15 real parameters are reduced to actually 5.
\[
\hat{\rho}^{(1,2)}=(1/4)\Big{(} \hat{I}^{\left( 1\right) }\otimes \hat{I}
^{\left( 2\right) }+\xi _{3}^{\left( 1\right) }\hat{\sigma}_{3}^{\left(
1\right) }\otimes \hat{I}^{\left( 2\right) }
\]
\begin{equation}
+\xi _{3}^{\left( 2\right) }\hat{I}^{\left( 1\right)
}\otimes \hat{\sigma}_{3}^{\left( 2\right)
}+\sum_{i}\zeta _{ii}\hat{\sigma}_{i}^{\left( 1\right) }\otimes \hat{\sigma}
_{i}^{\left( 2\right) }\Big{)},   \label{eq:(11)}
\end{equation}%
where the 5 parameters satisfy inequalities (6) and (7)
\begin{equation}
-1 \leq \zeta _{11,22}\leq 1,
\end{equation}
\begin{equation}
| \xi _{3}^{(1)}+\xi _{3}^{(2)} | -1 \leq \zeta _{33}\leq |\xi _{3}^{( 1)} -\xi _{3}^{(2)}| +1,
\end{equation}
\begin{equation}
-1 \leq \xi _{3}^{( 1)},\xi _{3}^{( 2)}\leq 1,
\end{equation}
and
\begin{equation}
{\xi} _{3}^{(1) 2} + {\xi} _{3}^{(2) 2}  +\sum_{i}\zeta_{ii}^{2} \leq 3.
\end{equation}
The density matrix (18) describes interesting polarization states of the pair of photons, that is, from the completely unpolarized state to that of the pure polarized state as defined in \cite{vem} and represents entangled photons provided (5) does not hold true for at least one of the $\zeta_{ij}$. All three examples that follow represent strictly nonseparable density matrices.

The eigenvalue equation (14) factors into the product of two quadratic polynomials and the eigenvalues are given by
\begin{eqnarray}
\lambda _{1,2} &=&\frac{1}{4}(1+\zeta _{33}\pm x_{+})  \label{eq:(13.a)} \\
\lambda _{3,4} &=&\frac{1}{4}(1-\zeta _{33}\pm x_{-}) , \notag
\end{eqnarray}%
where%
\begin{equation}
x_{\nu }=[\left( \xi _{3}^{\left( 1\right) }+\nu \xi _{3}^{\left( 2\right)
}\right) ^{2}+\left( \zeta _{11}-\nu \zeta _{22}\right) ^{2}]^{1/2},\text{ }%
(\nu =\pm ).  \label{eq:(13.b)}
\end{equation}

One obtains for the entropy (17) the expression
\[
S=S_{\max } -\frac{1}{4}\sum\limits_{\nu \text{ }=\text{ }\pm }\ln [\left(1+\nu \zeta _{33}+x_{\nu }\right) ^{1+\nu \zeta _{33}+x_{\nu }}
\]
\begin{equation}
\times\left( 1+\nu
\zeta _{33}-x_{\nu }\right) ^{1+\nu \zeta _{33}-x_{\nu }}] . \label{eq:(14)}
\end{equation}%
Entropy (25) depends on the three quantities $\zeta_{33}$, $x_{-}$, and $x_{+}$. The requirement that the eigenvalues (23) be real, positive quantities in order to give rise to a real valued entropy implies that
\begin{equation}
-1\leq \zeta_{33}\leq 1
\end{equation}
and
\begin{equation}
-1\mp\zeta_{33} \leq x_{\pm}\leq 1\pm \zeta_{33}.
\end{equation}
Note that the general expression for entropy (25) is a function of the three variables $\zeta_{33}$, $x_{-}$, and  $x_{+}$. For definiteness, we consider $\zeta_{33}\geq 0$. Entropy (25) assumes its maximum value $2\ln2$ when
\begin{equation}
\zeta_{33}=x_{-}=x_{+} =0
\end{equation}
(completely unpolarized state) and it assumes the value of zero (pure state of two photon polarization) when
\begin{equation}
\zeta_{33}=1, x_{-}=0, x_{+}=2.
\end{equation}
The requirement of nonnegative values for the eigenvalues $\lambda_{1},\cdots\lambda_{4}$ is also satisfied provided
\begin{equation}
1\leq p/2 +\zeta_{33}^2,
\end{equation}
\begin{equation}
|q|\leq p/2+\zeta_{33}^2-1,
\end{equation}
where $q$ is defined by
\begin{equation}
q=\xi{_3}^{(1)}\xi{_3}^{(2)}-\zeta_{33}-\zeta_{11}\zeta_{22}.
\end{equation}
The parameter $q$ also describes the purity of the two photon polarization state given by the density matrix (18), which is equal to zero for pure polarized states of the pair of photons \cite{vem}.

\subsection{Non exotic states in two-photon correlations}

Consider the correlated density matrix%
\[
\hat{\rho}^{(1,2)}=(1/4)\Big{(} \hat{I}^{\left( 1\right) }\otimes \hat{I}%
^{\left( 2\right) }+\zeta [ \hat{\sigma}_{1}^{\left( 1\right) }\otimes
\hat{\sigma}_{1}^{\left( 2\right) }-\hat{\sigma}_{2}^{\left( 1\right)
}\otimes \hat{\sigma}_{2}^{\left( 2\right) }]
\]
\begin{equation}
+\zeta_{33}\hat{\sigma}_{3}^{\left(
1\right) }\otimes \hat{\sigma}_{3}^{\left( 2\right)} \Big{)},
\label{eq:(16.a)}
\end{equation}%
where  the positivity of the eigenvalues requires that
\begin{equation}
-1\leq \zeta_{33} \leq 1
\end{equation}
and
\begin{equation}
-(1+\zeta_{33})/2 \leq \zeta \leq (1+\zeta_{33})/2.
\end{equation}
The von Neumann entropy (25) is
\[
S(\zeta,\zeta_{33})= S_{max}-\frac{1}{2}\ln(1-\zeta_{33})^{1-\zeta_{33}}
\]
\begin{equation}
-\frac{1}{4}\ln[(1+\zeta_{33}+2\zeta)^{1+\zeta_{33}+2\zeta}(1+\zeta_{33}-2\zeta)^{1+\zeta_{33}-2\zeta} ].
\end{equation}

The plot of the entropy (36) is shown in Fig. 1 where it is evaluated for different values of $\zeta_{33}$. Note that the range of values for $\zeta$ is restricted by the reality condition for the von Neumann entropy, viz., $0\leq \zeta\leq (1+\zeta_{33})/2$. The decrease of the entropy for increasing values of $\zeta_{33}$, for fixed values of $\zeta$, is to be expected since increasing the value of $\zeta_{33}$ means that the state of the two photons becomes less chaotic and approaches the entropy associated with the reduced entropy (12).

\begin{figure}
\begin{center}
   \includegraphics[scale=0.3]{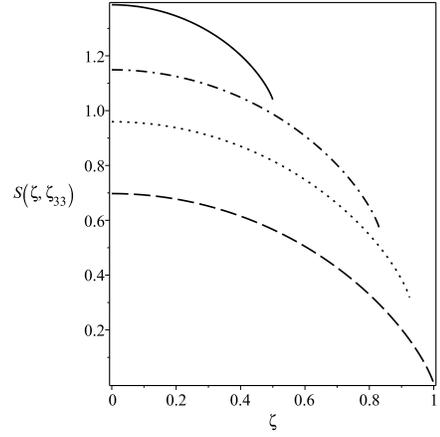}
\end{center}
\label{fig.theFig}
  \caption{  Plot of the entropy (36) for $\xi_{3}^{(1)}=\xi_{3}^{(2)}=0$, $\zeta_{11}=-\zeta_{22}=\zeta$, which gives $x_{+}=2|\zeta|$ and $x_{-}=0$. Inequality (27) becomes $0\leq \zeta_{33}\leq 1$ and $0\leq \zeta\leq (1+\zeta_{33})/2$. The values for $\zeta_{33}$ are as follows: $\zeta_{33}=0$ (solid), $\zeta_{33}=0.66$ (dashdot), $\zeta_{33}=0.85$ (dot), and $\zeta_{33}=1.0$ (single photon entropy (12))(dash).}
\end{figure}

Note that from the separability condition (3), it does not follow that states that satisfy $S\geq S_{1}$ are necessarily separable, In fact, in Fig. 1, we have non-exotic ($S>S_{1}$) entangled states.

\subsection{Exotic to non-exotic transition in single- and two-photon correlations}

The previous example and the one to follow below given by (40) are in agreement with our common understanding of entropy as a measure of the disorder of a system. In both these cases, the entropy is a monotonically decreasing function of the polarization. However, the system in this following second example gives rise to a very different behavior of the entropy as a function of polarization.

Suppose that two of the eigenvalues in (23) are set equal to zero, viz., $\lambda_{2}=\lambda_{4}=0$, and with $\zeta_{11}=\zeta_{22}=\zeta$, then the density matrix (18) becomes
\[
\hat{\rho}^{(1,2)}=(1/4)\Big{(} \hat{I}^{(1)}\otimes \hat{I}^{( 2)}+\xi _{3}^{( 1)}\hat{\sigma}_{3}^{(1) }\otimes \hat{I}^{(2)}+\xi _{3}^{(2)}\hat{I}^{(1)}\otimes \hat{\sigma}_{3}^{(2)}
\]
\begin{equation}
+\sqrt{(1-\xi _{3}^{( 1)})(1-\xi _{3}^{( 2)})}[\hat{\sigma}_{1}^{( 1)}\otimes \hat{\sigma}_{1}^{( 2)} +\hat{\sigma}_{2}^{( 1)}\otimes \hat{\sigma}
_{2}^{( 2)} ]
\end{equation}
\[
+(\xi _{3}^{( 1)}+\xi _{3}^{( 2)}-1)\hat{\sigma}_{3}^{( 1)}\otimes \hat{\sigma}_{3}^{( 2)}\Big{)}.
\]

Entropy (25) for the density matrix (37) is
\[
S(\xi_{3}^{( 1)},\xi_{3}^{( 2)}) = - \frac{\xi_{3}^{( 1)}+\xi_{3}^{( 2)}}{2} \ln(\frac{\xi_{3}^{( 1)}+\xi_{3}^{( 2)}}{2})
\]
\begin{equation}
- (1-\frac{\xi_{3}^{( 1)}+\xi_{3}^{( 2)}}{2}) \ln(1-\frac{\xi_{3}^{( 1)}+\xi_{3}^{( 2)}}{2})=S_{1}(\xi_{3}^{(1)}+\xi_{3}^{(2)}).
\end{equation}
Note that the dependence of entropy (38) on the Stokes parameters $\xi_{3}^{( 1)}$ and $\xi_{3}^{( 2)}$ is via their sum while such is not the case for the density matrix (37).

The case $\xi_{3}^{( 1)}=\xi_{3}^{( 2)}=\xi$, shown in Fig. 2 by the solid plot, gives for the entropy (38)
\begin{equation}
S(\xi)= -\xi\ln(\xi) -(1-\xi)\ln(1-\xi),
\end{equation}
which is actually the binary entropy function $h(\xi)$ introduced by Wootters \cite{Woot} in the definition of the entanglement of formation (EOF). Also, one obtains the reduced single photon entropy (12), dash plot in Fig. 2, for $\xi_{3}^{(1)}+\xi_{3}^{( 2)}=1-\xi$.

\begin{figure}
\begin{center}
   \includegraphics[scale=0.3]{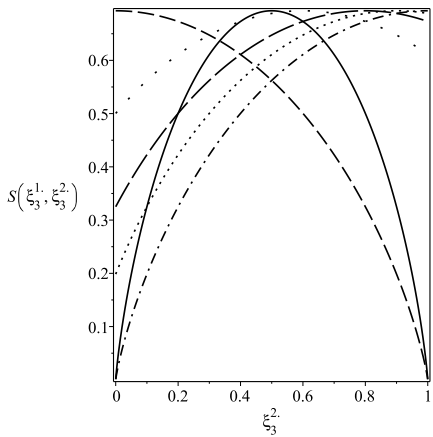}
\end{center}
\label{fig:theFig}
  \caption{Plot of the two photon entropy (38) as a function of $\xi_{3}^{( 1)}$ for different values of $\xi_{3}^{( 2)}$. The values are as follows: $\xi_{3}^{(1)}=\xi_{3}^{(2)}$ (solid), $\xi_{3}^{(2)}=0$ (dashdot), $\xi_{3}^{(2)}=0.1$  (dot), $\xi_{3}^{(2)}=0.2$ (longdash), $\xi_{3}^{(2)}=0.4$ (spacedot), $\xi_{3}^{(2)}=1-\xi-\xi_{3}^{( 1)}$ (dash). The dash plot corresponds to the single photon entropy (12) which intersects the other graphs at $\xi_{3}^{( 1)} =(1-\xi_{3}^{( 2)})/2$. }
\end{figure}

Fig. 2 shows the reduced single photo entropy (12), dash graph, together with the two photon entropy (38) as a function of the Stokes parameter for photon 1, $\xi_{3}^{( 1)}$, for various fixed values of the Stokes parameter for photon 2, $0\leq \xi_{3}^{( 2)}<1 $. The points of intersection between the dash plot and the other plots represent the transition points whereby  $S(\xi_{3}^{( 1)}, \xi_{3}^{( 2)})$, as a function of $\xi_{3}^{( 1)}$, goes from the region where $ S_{1}> S$ (exotic states) to the region where $S>S_{1}$ (non-exotic states), that is, where the quantum conditional entropy (1) changes sign from negative to positive.

\subsection{Exotic states in single- and two-photon correlations}

\begin{figure}
\begin{center}
   \includegraphics[scale=0.3]{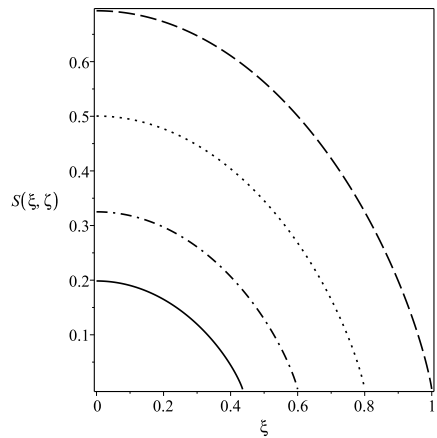}
\end{center}
\label{fig:theFig}
  \caption{Plot of the entropy (41) for $\xi_{3}^{(1)}=\xi_{3}^{(2)}=\xi$, $\zeta_{11}=-\zeta_{22}=\zeta$ and $\zeta_{33}=1$ , which gives $x_{+}=2 \sqrt{\xi^2+ \zeta^2}\leq 2$ and $x_{-}=0$. The values for $\zeta$ are as follows: $\zeta=0$  (single photon entropy (12))(dash), $\zeta=0.6$ (dot), $\zeta=0.8$ (dashdot), and $\zeta=0.9$ (solid). }
\end{figure}

In order to the find the role of the polarization states of the individual photons in the entropy of the two-photon system, we consider the case where the Stokes parameters  $\xi_{3}^{(1)}= \xi_{3}^{(2)}\equiv\xi \neq 0$ and the photon correlation parameters $\zeta_{11}= -\zeta_{22}\equiv \zeta$, and $\zeta_{33}=1$. The latter choice of parameters will allow the pure state to be realized when $\zeta=0$ for $\xi=1$ in the entropy (41) given below. The corresponding density matrix is
\[
\hat{\rho}^{1,2}=(1/4)\Big{(} \hat{I}^{(1)}\otimes \hat{I}^{(2)}+\xi [ \hat{\sigma}_{3}^{(1)}\otimes\hat{I}^{(2)}+\hat{I}^{\left( 1\right) }\otimes \hat{\sigma}_{3}^{\left( 2\right) }]
\]
\begin{equation}
 +\zeta[\hat{\sigma}_{1}^{(1)}\otimes \hat{\sigma}_{1}^{(2)} - \hat{\sigma}_{2}^{(1)}\otimes \hat{\sigma}_{2}^{(2)}]  + \hat{\sigma}_{3}^{\left( 1\right) }\otimes
\hat{\sigma}_{3}^{\left( 2\right) }\Big{)}.
\end{equation}

The entropy is
\begin{equation}
S(\xi,\zeta)=S_{\max }-\frac{1}{4}\ln \left[( 2+x_{+} \right) ^{2+x_{+} }\left( 2-x_{+}
\right) ^{2-x_{+} }] , \label{eq:(17.b)}
\end{equation}
where
\begin{equation}
x_{+}=2\sqrt{\xi^2+\zeta^2}\leq 2,
\end{equation}
and
\begin{equation}
0\leq \xi\leq 1.
\end{equation}
As can be seen in Fig. 3, the two-photon entropy $S(\xi,\zeta)$ mimics that of the single photon entropy $S_{1}(\xi)$, that is, the single photon entropy (12) decreases with increasing values of $\xi$, while the two photon entropy (41) decreases for given value of $\zeta$ with increasing values of $\xi$ or conversely.

These results indicate that once again one has exotic, entangled states since $S_{1}(\xi) >S(\xi,\zeta)$ for $0\leq \xi \leq \sqrt{1-\zeta^2}$, that is, the conditional entropy is negative for these values of the parameters.

\section{Entropy and measurement}

In this section, we consider the effect of photon polarization measurements with the aid of an efficiency matrix and the resulting reduction of the two-photon density matrix. Of course, the change of the entropy of the two-photon system depends crucially on the particular measurement that is actually performed \cite{AP95}.

Three types of polarization measurements can be carried out on the two-photon system.\\

\textbf{I}. A standard measurement of two independent analyzers is performed with the aid of the efficiency matrix of the polarization filter \cite{KB12} given by the direct product of the Stokes matrices of the analyzers $\hat{\epsilon}_{I} =\hat{\rho}_{1}\otimes \hat{\rho}_{2}$, where  $\hat{\rho}_{i}=\frac{1}{2}(\hat{I}^{(i)}+ \boldsymbol{n}_{i}\cdot \hat{\boldsymbol{\sigma}}^{(i)})$ and $\boldsymbol{n}_{i}$ is a unit vector ($i=1,2$). After the measurement, one has a separable polarization state of the pair of photons. The probability of reduction or correlation parameter is then
\begin{equation}
w_{I}= \textup{Tr}\big{(} \hat{\epsilon}_{I} \hat{\rho}^{(1,2)} \Big{)}= \frac{1}{4}\big{(}1+\boldsymbol{n}_{1}\cdot\boldsymbol{\xi}^{(1)}+\boldsymbol{n}_{2}\cdot\boldsymbol{\xi}^{(2)}+\sum_{i,j}  \zeta_{ij}n_{1i}n_{2j}\big{)}.
\end{equation}

The probability $w_{I}$ is actually less than 1, owing to inequalities (6) and (7), which means that this type of measurement disturbs the system and reduction of the initial quantum state, either a pure entangled or mixed state, takes place into a pure polarization state and so the total entropy is zero. That is to say, the entropy of the system decreases owing to the measurement.\\

\textbf{II}. Measurement with a two-photon analyzer where the efficiency matrix of the polarization filter is coincident with the density matrix (4) with 15 filter parameters for a pure polarization state, viz., $\boldsymbol{\xi}_{1}$, $\boldsymbol{\xi}_{2}$, and  $\zeta _{ij}^{f}$,
\[
\hat{\epsilon}_{II}=(1/4)\Big{(} \hat{I}^{( 1)}\otimes \hat{I}^{( 2) }+ \boldsymbol{\xi}_{1}\cdot\hat{\boldsymbol{\sigma}}^{(1)}\otimes \hat{I}^{( 2) }
\]
\begin{equation}
+\hat{I}^{( 1)}\otimes \boldsymbol{\xi}_{2}\cdot\hat{\boldsymbol{\sigma}}^{(2)}+\sum_{i,j}\zeta _{ij}^{f}\hat{\sigma}_{i}^{( 1)}\otimes \hat{\sigma}_{j}^{( 2)}\Big{)}.
\end{equation}
This two-photon analyzer reduces the initial two photon polarization state to that of a pure polarization state for the two photon system determined by the 15 filter parameters. One obtains for the probability $w_{II}$ of reducing the initial state to the filtered state
\begin{equation}
w_{II}= \textup{Tr}\big{(} \hat{\epsilon}_{II} \hat{\rho}^{(1,2)} \Big{)} = \frac{1}{4}\big{(}1+\boldsymbol{n}_{1}\cdot\boldsymbol{\xi}^{(1)}+\boldsymbol{n}_{2}\cdot\boldsymbol{\xi}^{(2)}+\sum_{i,j}\zeta_{ij}^{f}\zeta_{ij}  \big{)}.
\end{equation}
In contrast to the probability $w_{I}$, which is always less than one, probability $w_{II}$ could even reach the value of one. If the initial two photon state is in a pure polarization state, then the measurement does not disturbs the system and so the entropy does not change and remains equal to zero. If, however, the initial two-photon state is in a mixed state, then the measurement actually forces the system from an initial finite value of the entropy to zero entropy. Accordingly, as was the case in the previous measurement, this measurement also has the effect of reducing the entropy of the two-photon system after the polarization of the system has been measured.\\

\textbf{III}. This third measurement can be performed by one analyzer with efficiency matrix of the polarization filter $ \hat{\epsilon}_{III}$ given by
\begin{equation}
 \hat{\epsilon}_{III}= \frac{1}{2}(\hat{I}^{(1)}+ \boldsymbol{n}\cdot \hat{\boldsymbol{\sigma}}^{(1)}) \otimes \hat{I}.
\end{equation}
Photon 2 get polarized by this measurement on photon 1 and photon 2 is now described by the Stokes matrix \cite{vem}
\begin{equation}
\hat{\rho}_{2}^{'} =\textup{Tr}\big{(} \hat{\epsilon}_{III} \hat{\rho}^{(1,2)} \Big{)}
\end{equation}
with Stokes parameters
\begin{equation}
\xi_{j}^{(2)'} =\Big{[}\xi_{j}^{(2)}+ \sum_{i}\zeta_{ij} n_{i}\Big{]}\Big{(}1+\boldsymbol{n}\cdot \boldsymbol{\xi}^{(1)}\Big{)}^{-1}.
\end{equation}
Therefore, with the aid of the expression for the entropy for a single photon, the entropy of photon 2 becomes after the measurement of photon 1
\begin{equation}
S(\boldsymbol{n})=\ln(2) -\frac{1}{2}\ln \big{[}(1-\xi^{(2)'})^{1-\xi^{(2)'}}(1+\xi^{(2)'})^{1+\xi^{(2)'}} \big{]},
\end{equation}
where the general expression for the magnitude of the Stokes vector $\boldsymbol{\xi}^{(2)'}$ after the measurement is
\[
\xi^{(2)'} =\sqrt{(\xi_{1}^{(2)'})^2  +(\xi_{2}^{(2)'})^2 +(\xi_{3}^{(2)'})^2  }
\]
\begin{equation}
=\frac{\sqrt{n_{1}^2\zeta_{11}^2  + n_{2}^2 \zeta_{22}^2  +(\xi_{3}^{(2)}+n_{3}\zeta_{33})^2     }  } {1+n_{3}\xi_{3}^{(1)}},
\end{equation}
where the second expression in (51) follows for the 5-parameter density matrix (18).

The analyzer $\hat{\epsilon}_{III}$ depends on the unit vector \textbf{n} and the final state of photon 2 depends also on the initial values of the Stokes parameters of photons 1 and 2. In what follows, we show that depending on the vector \textbf{n} and the initial states of photons 1 and 2, the effect of the measurement on photon 1 will result in the increase or decrease of the entropy of photon 2.

\subsection{Entropy of reduced density with only two-photon correlations}

The correlated density matrix (33) yields for the magnitude of the Stokes vector (51)
\begin{equation}
\xi^{(2)'} = \sqrt{(1-n_{3}^2)\zeta^2 +n_{3}^2\zeta_{33}^2 }.
\end{equation}
Notice that (52) is an even function of $n_{3}$. Fig. 4 shows the behavior of the reduced entropy $S(\zeta,\zeta_{33}, n_{3})$ of particle 2 for various values of $\zeta_{33}$ and $n_{3}$. There is both ``cooling" $(S<S_{1})$ and ``heating" $(S>S_{1})$ of photon 2 after the measurement of photon 1.
\begin{figure}
\begin{center}
   \includegraphics[scale=0.3]{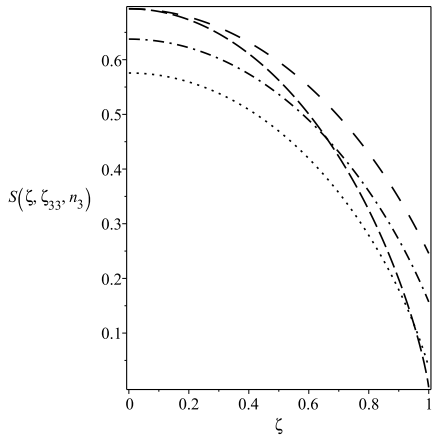}
\end{center}
\label{fig:theFig}
  \caption{Plot of the photon reduced entropy (50) with magnitude (52) for $\xi^{(2)'}$. The values for $\zeta_{33}$ and $n_{3}$ are as follows: $\zeta_{33} =0$, $n_{3}=0.5$ (spacedash), single photon entropy (12) (dash), $\zeta_{33} =0.66$, $n_{3}=0.5$ (dashdot), and $\zeta_{33} =0.95$, $n_{3}=0.5$ (dot).}
\end{figure}

\subsection{Entropy of reduced density with unequal single correlation and two-photon correlations}

The correlated density matrix (37) yields for the magnitude of the Stokes vector (51)
\begin{equation}
\xi^{(2)'} = \frac{1}{ 1+n_{3}\xi_{3}^{(1)} }
\end{equation}
\[\times \sqrt{(1-n_{3}^2)(1-\xi_{3}^{(1)})(1-\xi_{3}^{(2)})
+\Big{[}n_{3}(\xi_{3}^{(1)}+\xi_{3}^{(2)} -1) +\xi_{3}^{(2)}\Big{ ]}^2}       .
\]
Note that (53) is not an even function of $n_{3}$ and so negative values may be considered. Fig. 5 shows the behavior of the reduced entropy $S(\xi,n_{3})$ for $\xi_{3}^{(1)}= \xi_{3}^{(2)}=\xi$ as a function of $\xi$ for given values of $n_{3}$. In this case, the entropy of photon 2, after measurement of photon 1, exhibits ``cooling" and ``heating" of the unmeasured photon 2.  Fig 6 shows the case where $n_ {3} $ assumes a negative value and shows considerable ``cooling" of the unmeasured photon 2.

\begin{figure}
\begin{center}
   \includegraphics[scale=0.3]{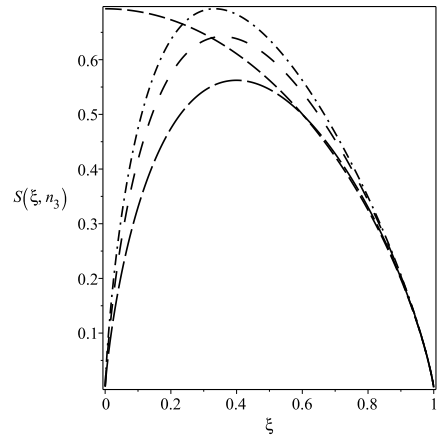}
\end{center}
\label{fig:theFig}
  \caption{Plot of the photon reduced entropy (50) with magnitude (53) for $\xi^{(2)'}$. The values of the Stokes parameters are set as follows $\xi_{3}^{(1)}=\xi_{3}^{(2)}=\xi$. The values for $n_{3}$ are as follows: $n_{3}=1$ (dashdot), $n_{3}=0.8$ (spacedash), $n_{3}=0.5$ (longdash), and the single photon entropy (12) (dash).}
\end{figure}

\begin{figure}
\begin{center}
   \includegraphics[scale=0.3]{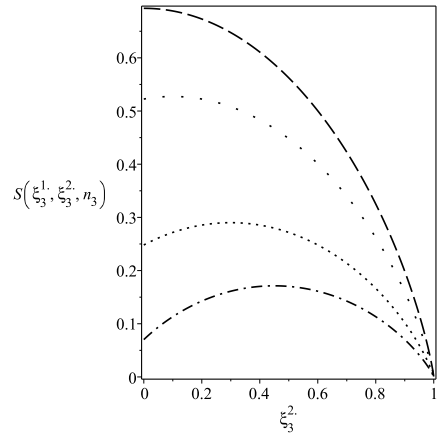}
\end{center}
\label{fig:theFig}
  \caption{Plot of the photon reduced entropy (50) with magnitude (53) for $\xi^{(2)'}$. The values for $\xi_{3}^{(1)}$ and $n_{3}$ are as follows: $\xi_{3}^{(1)} =0.1$, $n_{3}=-0.3$ (dashdot), single photon entropy (12) (dash), and $\xi_{3}^{(1)} =0.4$, $n_{3}=-0.3$ (dot), and $\xi_{3}^{(1)} =0.8$, $n_{3}=-0.3$ (spacedot).}
\end{figure}

\begin{figure}
\begin{center}
   \includegraphics[scale=0.3]{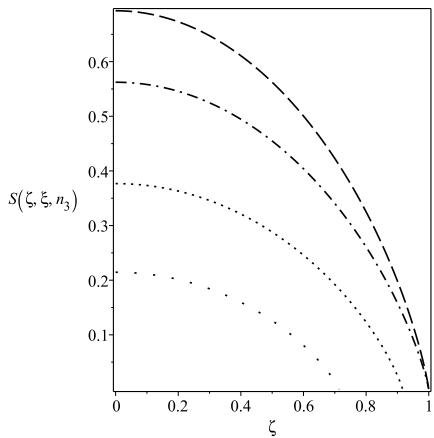}
\end{center}
\label{fig:theFig}
  \caption{Plot of the photon reduced entropy (50) with magnitude (54) for $\xi^{(2)'}$. The values for $\xi$ and $n_{3}$ are as follows: $\xi =0$, $n_{3}=0.5$ (dashdot), single photon entropy (12) (dash), $\xi =0.4$, $n_{3}=0.5$ (dot), and $\xi =0.7$, $n_{3}=0.5$ (spacedot).}
\end{figure}

\subsection{Entropy of reduced density with equal single correlation and two-photon correlations}
The correlated density matrix (40) yields for the magnitude of the Stokes vector (51)
\begin{equation}
\xi^{(2)'} = \frac{\sqrt{(1-n_{3}^2) \zeta^2 + (\xi+n_{3})^2 }}{ 1+n_{3}\xi}
\end{equation}
Note that (54) is not an even function of $n_{3}$ and so negative values may be considered. Fig. 7 shows the behavior of the reduced entropy $S(\zeta, \xi, n_{3})$ as a function of $\zeta$ for given values of $\xi$ and $n_{3}$. Fig. 8 shows the case where $n_{3}$ assumes a negative value. In both instances, we have only ``cooling."

\begin{figure}
\begin{center}
   \includegraphics[scale=0.3]{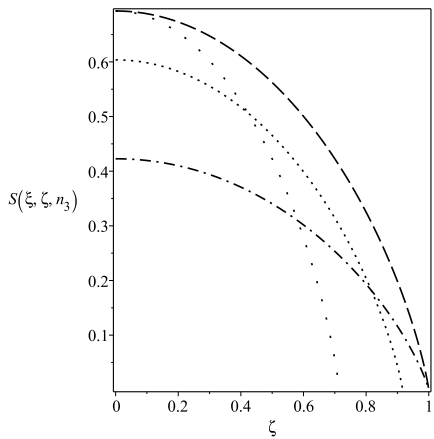}
\end{center}
\label{fig:theFig}
  \caption{Plot of the photon reduced entropy (50) with magnitude (54) for $\xi^{(2)'}$. The values for $\xi$ and $n_{3}$ are as follows: $\xi =0$, $n_{3}=-0.7$ (dashdot), single photon entropy (12) (dash), $\xi =0.4$, $n_{3}=-0.7$ (dot), and $\xi =0.7$, $n_{3}=-0.7$ (spacedot).}
\end{figure}

\section{Summary and discussion}
We have obtained exact results for the joint von Neumann entropy for the polarization states of a two-photon system governed by a 5-parameter polarization density matrix and studied the sign of the quantum conditional entropy. We find that the quantum conditional entropy may assume positive or negative values. The latter indicates the presence of exotic states. We have also studied the resulting reduced density matrix of one of the photons after a measurement is performed on the second photon. We find a sort of ``heating" and ``cooling" of the unmeasured photon where the final entropy of the unmeasured photon has either increased or decreased. We believe that these results may be of interest in the general area of quantum computation and quantum information theories.

\emph{}
\end{document}